# Condensation of earthquake location distributions: optimal spatial information encoding and application to multifractal analysis of South Californian seismicity

Y. Kamer[1, 3], G. Ouillon[2], D. Sornette[3], J. Wössner[4]

[1]Swiss Seismological Service, ETH Zürich, Switzerland
[2]Lithophyse, 4 rue de l'Ancien Sénat, 06300 Nice, France
[3]Department of Management, Technology and Economics, ETH Zürich, Switzerland
[4]Risk Management Solutions Inc., Stampfenbachstrasse 85, Zürich 8006, Switzerland

**Abstract**: We present the "condensation" method that exploits the heterogeneity of the probability distribution functions (PDF) of event locations to improve the spatial information content of seismic catalogs. As its name indicates, the condensation method reduces the size of seismic catalogs while improving the access to the spatial information content of seismic catalogs. The PDFs of events are first ranked by decreasing location errors and then successively condensed onto better located and lower variance event PDFs. The obtained condensed catalog differs from the initial catalog by attributing different weights to each event, the set of weights providing an optimal spatial representation with respect to the spatially varying location capability of the seismic network. Synthetic tests on fractal distributions perturbed with realistic location errors show that condensation improves spatial information content of the original catalog, which is quantified by the likelihood gain per event. Applied to Southern California seismicity, the new condensed catalog highlights major mapped fault traces and reveals possible additional structures while reducing the catalog length by ~25%. The condensation method allows us to account for location error information within a point based spatial analysis. We demonstrate this by comparing the multifractal properties of the condensed catalog locations with those of the original catalog. We evidence different spatial scaling regimes characterized by distinct multifractal spectra and separated by transition scales. We interpret the upper scale as to agree with the thickness of the brittle crust, while the lower scale (2.5km) might depend on the relocation procedure. Accounting for these new results, the Epidemic Type Aftershock Model formulation suggests that, contrary to previous studies, large earthquakes dominate the earthquake triggering process. This implies that the limited capability of detecting small magnitude events cannot be used to argue that earthquakes are unpredictable in general.

## 1- Introduction

The latest advances in the instrumentation field have increased the station coverage and lowered seismic event detection thresholds. This has resulted in a vast increase in the yearly number of located events. The abundance of data comes as a double-edged sword: while it facilitates more robust statistics, this comes at the cost of larger computations, with execution times often growing exponentially with the number of data points. In many analyses studying temporal or spatial clustering, a common approach to deal with the large amount of data is to introduce threshold criteria. These can be minimum magnitude, maximum location uncertainty or a specific time or space window. While large magnitude events are of greater importance for risk assessment studies, events with small location uncertainties are crucial for accurately mapping the active/potential unknown structures [1,2]. Some studies prefer limiting their data to the most recent periods, however, there is strong evidence that the notion of time invariance does not hold for seismicity distributions, at least on the time scales covered by existing catalogs. Apart from containing a degree of arbitrariness, implementing these thresholds discards data that clearly contain some information, and could potentially be useful.

In this study, we present a novel method that (i) assesses the relative importance of each earthquake hypocenter location data point using its uncertainty as a metric, (ii) reduces the size of the dataset, (iii) preserves the total number of events and (iv) helps improving the signal to noise ratio for statistical analyses such as the multifractal analysis of the spatial distribution of hypocenters. Our motivation stems from the fact that the geometrical information contained in a seismic catalog is not optimally encoded, is redundant and thus requires unnecessary memory. In other words, the same spatial information can be stored allocating fewer resources. This encoding inefficiency is a result of the sequential data entry where new events are recorded without taking into account the information contained in previous ones. As an extreme example, consider several events with identical parameters occurring at the same location. For the purpose of spatial clustering, it would be optimal to group together and represent them by a single entry with a multiplier (weight). Instead, they each occupy a memory space as if they provided distinct spatial information. We can generalize this idea for events with locations that are not identical but relatively close: given its mean vector and covariance matrix of position errors, each event can be regarded as a representation of the distribution of its possible locations (i.e. a probability density function, PDF). We propose to implement the re-grouping at this micro-scale in order to optimize the encoding of the joint spatial distribution of all events in a catalog. In this way, we also minimize biases of statistical estimators for variables depending on the whole spatial information.

The method does not rely on any assumptions regarding the physical processes generating the point distribution. As such, the method can be readily applied to other point process datasets featuring location uncertainty (e.g. meteorite impacts, explosions, sunspots…). To facilitate a non-arbitrary implementation, the following physical analogy is useful: if each event entry describes the possible location distribution of the event (i.e. microstates), the logarithm of the squared deviations can be regarded as its entropy. The higher the entropy is, the wider is the scatter of microstates. With this definition in mind, the condensation phenomena can be viewed as water vapor (high entropy state) liquefying upon encountering a low entropy setting (lower temperature). Following the same analogy, we can imagine events with high location uncertainties as clouds of water vapor and the ones with more certain locations as droplets of cold water. Under such a setting, the vapor would condense onto the nearest droplets reducing the overall volume (i.e data size) while conserving the total weight content. It is important to note that in this thermodynamic analogy the entropy is related to the actual scatter/disorder of the microstates of gases and liquids. In the case of earthquake locations, the entropy is an expression of the lack of precise knowledge due to observational limitations.

Similar concepts of spatial clustering and entropy have been used in previous studies [3,4], with the main difference that these aimed at relocation of events to reveal underlying structures. The collapsing method introduced by Jones and Stewart [3] iteratively moves each event towards the centroid of all events enclosed by its uncertainty ellipsoid. The process stops when the distribution of all movements becomes comparable to the variance of the spatial distribution of the catalog. This method successfully highlights structures by collapsing multiple events onto each other. However, synthetic tests with uniformly distributed random points produced similar linear structures that could be proven as artifacts of the method. To counter this downside, Nicholson et al. [4] introduced a weighting function in the calculation of the centroid. However, their synthetic test showed that the artifacts were still present, only reduced in amplitude. The method presented in this paper 1) preserves the total seismicity rate of a catalog while minimizing the data size without altering the location of the events, 2) illuminates possible structures as well-located events are highlighted by the condensation of the poorly located ones in their vicinity and 3) regularizes the catalogs with respect to the spatially varying location quality. This provides an efficient encoding of the location quality for further analysis such as inter-event distances.

The paper is structured as follows. First, we present an illustrative description of the condensation method and its application to Southern Californian seismicity. In Section 3, we investigate quantitatively and visually the effect of condensation in terms of likelihood gain and weight transfer. In Section 4, we perform a comparative multifractal analysis on the original and condensed catalogs. We conclude our paper with a discussion of our findings and their implications for interpreting previous studies and for future studies.

## 2- The Condensation Method

### *2.1 Description and illustration of the method*

Condensation transfers weight from events with large location uncertainty (“sources”) to events with smaller location uncertainty (“targets”). For this purpose, reported horizontal and vertical standard location errors are used to estimate an isotropic overall variance for each event in the catalog (Equation 1).

$$Var_{iso} = {\sigma_x}^2 + {\sigma_y}^2 + {\sigma_z}^2 \qquad (1)$$

The implementation of the method follows the following steps:

1) Assign a unit weight to each event and sort them in descending order according to their overall variance. Select the event with the highest variance as a source.

2) Sample the source's location PDF using a large number of points (typically 1,000; labeled as offsprings).

3) Compute the responsibility of each event (targets and current source) in generating each offspring defined in 2). This is done by estimating the likelihood of the offspring conditioned on the target.

4) Each event gains a weight proportional to the ratio of the number of offspring for which it had the largest responsibility.

5) We consider the next source event and go to step 2.

The condensation stops when all events except the ones with the lowest variance have been processed, i.e., when there are no target events for the latter to condense on. For a better understanding, we present a step by step illustration of the method applied to a set of 1D normal distributions representing location of five events (Figure 1). The distributions have respectively variances of 0.5, 1, 1.5, 2, and 2.5 and are labeled with letters A to E. The event E, having the largest variance, becomes the initial source. Condensation continues until all events except A, which has the lowest variance, have transferred their weight to targets providing the higher likelihoods (Figure 1). At the end of the condensation, one observes that the weights of events E and C decrease while those of events A and B increase, the total sum of weights being preserved. The weight of event D remains almost unchanged since no event with better or worse variance is present in its vicinity. For applications to seismic catalogs, this indicates that, in areas with sparse seismicity, condensation will tend to conserve the initial weight distribution (i.e. the initial structure of seismicity). However, many catalogs feature pronounced spatial clustering of events, which may differ significantly in their location accuracies. Therefore, condensation leads to a large fraction of events ending up with vanishing weights. Discarding these events allows for a reduction in the overall data size of the catalog without any loss in the overall information on the spatial structure encoded in the catalog, as we demonstrate below.

The output of catalog condensation is a vector of weights assigned to each individual event in the original input catalog. The sum of these weights is equal to the total number of events in the input catalog. Events can be classified into the following four categories according to their final condensation weights:

a. *Weight >1*: Events that are better located with respect to their neighbors and have thus increased their weights. Note that good location accuracy does not necessarily lead to increased weights since this depends on the local presence of events with higher location uncertainty. Being selected as a source or a target depends on the respective accuracies of the events with overlapping location error PDFs.
b. *Weight <1*: Events that have relatively poor location accuracy and are in the close proximity of better located events. It is likely that these events will further decrease their weights in the future, as location capabilities of the seismic network improve and as new events with better accuracy are recorded in their vicinity.
c. *Weight =1*: This occurs for spatially isolated events that neither gained nor lost weight. This can also be observed when several events are close together and have identical variances. Another possibility is that during condensation, an event acquires the exact weight that it loses and hence ends up with a weight ≈1. However, due to computational precision, the final weight is unlikely to be precisely 1.

d. *Weight =0*: Events whose spatial PDF information can be virtually expressed as a combination of other better located events. These events can be discarded and hence reduce the catalog's length significantly.

The weights of the condensed catalog can be regarded as coefficients of spatial importance optimized with respect to the spatially varying location capability of the seismic network.

### *2.2 Synthetic Test with Fractal Distributions*

In this subsection, we generate synthetic fractal distributions, perturbed with realistic locations errors, and compare their spatial information content before and after condensation. We consider seven distributions (Figure 2), with different fractal dimensions *D*=[3.00, 2.58, 2.00, 1.58, 1.00, 0.63, 0] generated by recursive replication of template 3D density matrices (for a detailed description of this process the reader is referred to Figure 3 of [5]). The fractals are generated within a cube of 10km side length and contain 3360 points. For a realistic representation of location uncertainties, we use the covariance matrices of 3360 aftershock events following the 1992 M7.3 Landers earthquake [2]. The median values of the square roots of the eigenvalues are $\sigma_I$=0.89 km, $\sigma_{II}$=0.39 km and $\sigma_{III}$=0.27 km. The initial locations are stored as the true locations and then perturbed using the randomly assigned covariance matrices. The resulting perturbed locations and the covariance matrices become the perturbed catalog. This perturbed catalog is then condensed to obtain a vector of weights. We then calculate the likelihood of the true (unperturbed) locations with respect to the perturbed ($L_P$) and condensed catalog ($L_C$) according to the following formulation:

$$
\begin{aligned}
L_P &= \prod_{i=1}^{N} \sum_{j=1}^{N} \tfrac{1}{N} f(t_i, r_j, \hat{\sigma}_j) \\
L_C &= \prod_{i=1}^{N} \sum_{j=1}^{N} \tfrac{w_j}{N} f(t_i, r_j, \hat{\sigma}_j)
\end{aligned}
\qquad (2)
$$

where *N* is the number of points, *f* is the multivariate normal probability density function, $t_i$ is the true location of point *i*, $r_j$ is the perturbed location, $\sigma_j$ is the covariance matrix and $w_j$ is the condensation weight of point *j*. Thus, we consider each catalog as a Gaussian mixture model represented by a weighted sum of the multivariate normal distributions associated with each event [6]. We calculate the probability that the true locations were generated by either of these models. This formulation allows us to quantify the likelihood gain for any given arbitrary set of true locations. Figure 3a shows the log likelihood gain per point ($\log(L_c/L_p)/N$) for the seven fractal distributions as a function of the number of data-points. We observe significant likelihood gains for all cases except the uniform (*D*=3.00) case. The gain increases with both the number of samples and the degree of clustering (decreasing *D*), thus suggesting that it is governed by the minimum spacing ($\Delta d$) between the points. To investigate this hypothesis, we calculate analytically $\Delta d$ for each fractal distribution using the following formula:

$$\begin{aligned} \Delta d &= 10 L^{-\log(N)/\log(M)} km \\ D &= \log(M)/\log(L) \end{aligned} \quad (3)$$

where $L$ and $M$ are the unit length and unit mass of the template matrices given in Figure 2. As expected, the likelihood gain increases as $\Delta d$ decreases (Figure 3b). However the fact that the data do not collapse on a single curve suggests that the minimum spacing is not the sole controlling parameter. Spatial clustering of small spacings, is another factor, as well as the interplay between the local anisotropy of the events' locations and the one of the uncertainty ellipsoids.

The likelihood gains obtained through condensation can be understood in terms of information retrieval. Each time an earthquake is located, the true location is perturbed with an error vector due to instrumental (signal to noise ratio) and modeling (Earth structure) errors. For a single event, having multiple wave arrival time observations allows one to estimate the amplitude of this error vector (expressed as the standard error) but not its orientation. Thus, each time we locate an earthquake, this information is lost. Condensation facilitates the retrieval of this orientation by exploiting the mutual information of proximate events and giving preference to the more certain ones.

## 3- Application to South Californian Seismicity

### *3.1 Condensing the catalog*

As a first application of the method, we use the waveform relocated Southern California earthquake catalog of Hauksson et al. [7]. Most of the events in this catalog have been relocated using differential travel times and a 3D velocity model. Since condensation is based on the absolute location quality of all events, we consider the absolute location errors that are provided as one-sigma errors for the horizontal and vertical components. As a pre-filter to reduce the size of the catalog, we exclude events with horizontal or vertical error larger than 20 km, resulting in a total of 493,025 events. Keeping these events would not change our results as their large location errors implies that the condensation method would make them disappear by distributing their mass approximately uniformly to a large number of neighboring events. The 493,025 events are ranked by their descending isotropic variances. Since the one-sigma errors are reported with a 100m resolution, the isotropic variance distribution becomes discrete and results in a total of 5,651 distinct groups. The events in each group are then used in turn as sources, transferring weights to target events in lower variance groups. The condensation reduces the weights of 111,487 poorly located events to zero, while increasing correspondingly the weights of better located events in their vicinity. This corresponds to an overall data length reduction (i.e. compression) of 22.6%. An investigation of the temporal distribution of events with zero weights reveals that they are most numerous (≈10,900) in 1992, the year of the Landers earthquake, which triggered a large number of aftershocks. The year with the second largest number of events with zero weight is 1994 (Northridge earthquake) with ≈9,700 events (40% of all events recorded that year). This implies that many aftershocks of these two major earthquakes in Southern California have been recorded with rather poor location quality relative to the other events, notwithstanding the relocation procedure.

The probability density distributions of the vertical, horizontal and isotropic errors over all events belonging to the original and condensed catalogs are shown in Figure 4. The change in the distributions depicts the weight transfer occurring between events with high and low spatial variance governed by their proximity and relative location error distributions.

In the presented study the condensation method is applied to a set of hypocenter (i.e nucleation points), and yields a new catalog of nucleation points (which is a subset of the original one). In that respect there is no reason to take account of the sources' spatial extent. However, if one is interested in a representation of the overall strain field, the spatial extends of the sources would be important. The analysis of nucleation points can be regarded as a robust approach due to the fact that any statistical analysis of seismicity is dominated by the more numerous small magnitude events [8]. For instance, constraining the analysis to events smaller than M5 (rupture length of ~1.4km [9]) reduces the number of data points only by 107 ( ~0.02% of all events) and thus the results remain unaffected. Furthermore, the extended geometry of large earthquakes is always illuminated by nucleation points of small magnitude aftershocks.

### *3.2 Visualizing and quantifying weight transfer due to condensation*

The original and the condensed catalogs can be considered as Gaussian mixture models, as discussed in the previous section. Thus, both catalogs are spatial probability density functions (PDF) of the observed events. Comparing two PDFs in 1D can be readily achieved using distance measures such as the Kolmogorov or Anderson-Darling statistics [10]. However, for multidimensional distributions, the extension of these measures becomes problematic as the result becomes dependent on the ordering (left-right, up-down) of the data [11].

Since seismicity is distributed within a volume, visualizing the weight transfer occurring as a result of the condensation would require volumetric density plots and slice planes at various angles, which become difficult to interpret. For a simpler illustration, in this section, we project all events onto the surface and thus omit the vertical dimension. With this simplification, condensation is performed in 2D, relying only on the horizontal location errors. In this 2D setting, the compression rate increases to 41.1%. To estimate the probability density distributions, the spatial extend of the catalog is discretized using a 100x100 m grid resulting in a 8052 x 8341 matrix. The probability density functions for each catalog are estimated at each of these grid points by summing up the weighted contributions of the PDFs of all individual events. Notice that the two catalogs differ only in the weights assigned to the events (a constant of 1 for the original and varying weights for the condensed one). The total PDF of the original catalog is then subtracted from the total PDF of the condensed one. The total displaced weight is calculated by integrating the absolute differences divided by two, in order to account for the fact that transferred weight is double-counted as it is reported as negative at the source location and as positive at the target. At the initial resolution of 100 m, we calculate that 14.7% of the total weight has been displaced due to condensation. We note that the percentage of displaced mass and the compression level (41.1%) are not directly comparable because the former is coarse grained at the scale of the grid cells (100m) while the latter is estimated on a pointwise basis (i.e. corresponding to a cell size → 0).

The percentage of displaced weight includes all displacement occurring at scales larger than the grid resolution (100m). To depict the effect of condensation more accurately, we quantify weight displacement as a function of distance. For this purpose, we apply a rotationally symmetric Gaussian low-pass filter to the total PDFs of the two catalogs (original and condensed). By varying the bandwidth of the filter, we can effectively coarse-grain the PDFs, so that the measure is insensitive to weight displacement occurring at scales smaller than the filter bandwidth. We observe that the total displaced weight decreases significantly as the scale of interest is increased (Figure 5c). This means that most of the condensation occurs between neighboring events, as can be expected.

To put these results in perspective, we compare the displaced weight as a function of the filter bandwidth curve with the same curve obtained by subtracting two random PDFs spanning the same extent as the seismicity catalog (Figure 5). The random PDFs were generated by assigning each cell with a random number drawn from a uniform distribution [0 1] and then normalizing the whole matrix so that it integrates to one. The difference of two standard independent uniform random variables ($X_1$,$X_2$) follows the standard triangular distribution given as:

$$
\begin{aligned}
&Y = X_1 - X_2 \\
&f_Y(y) = \begin{cases} y+1 & -1 < y < 0 \\ 1-y & 0 \le y < 1 \end{cases}
\end{aligned}
\tag{4}
$$

As discussed above, the analysis of displaced weight considers half of the absolute differences, whose distribution is given as:

$$
f_Y(z) = 2(1-z) \quad 0 \le z < 1 \tag{5}
$$

Equation (5) represents a right-angled triangle and as such this distribution has its mean at $z$=1/3. Thus, for two random PDFs, the expected displaced weight at the original resolution is 33%, which is considerably larger than the 14.7% observed for the original and condensed catalogs of Southern California seismicity. For increasing spatial bandwidths, the two curves given in Figure 5c show that the weight transfer due to condensation is focused at small scales (~1km) and decreases more rapidly as a function of scale for the natural catalog than for a random displacement process. The limited scale of mass transfer is also in agreement with the distributions of horizontal location errors given in Figure 4. These results show that condensation is consistent with the overall location error distribution and preserves the overall spatial features.

Note that, for the random PDF, the displaced weight scales as the inverse of the coarse-grained scale (slope -1 in the log-log plot). This can be explained as follows. Choosing a coarse-grained scale $\lambda$ on an arbitrary location $r$ can be roughly represented as summing the weight within a disk of radius $\lambda$ centered on $r$ and asking how much of the weight within that disk is transferred outside that disk and how much weight within this disk comes from outside, via the operation of the condensation method. The transfer of weight operates through the overlapping of the PDFs of the sources and the PDFS of the targets. At coarse-graining scale $\lambda$, this translates into calculating the overlapping of the source PDFs outside the target disk centered on *r* with the PDFs inside the disk. In the limit where $\lambda$ is larger than the typical standard deviations of the PDFs, and when the positions of the events are uniformly random, this involves a weight proportional to the perimeter of the disk, i.e. proportional to $\lambda$. Relative to all the weight transferred between PDFs within the disk that remain within the disk, which is proportional to its surface $\lambda^2$ for a uniform random PDF, this gives $\lambda/\lambda^2$, which is the inverse scaling shown in Figure 5c for the random field. For the Southern California catalog, we observe that the displaced weight scales approximately as $1/\lambda^{1.5}$, with a slight downward curvature. Roughly speaking, the real catalog has less "perimeter" and more "surface" or bulk concentration, which is nothing but an expression of spatial clustering.

For a qualitative inspection, in Figure 6, we superimpose the fault traces obtained from the Community Fault Model [12] onto the map of weight depletion/enrichment coarse-grained at bandwidth $\sigma$=3km. This value is chosen in agreement with previous fault observations of apparent low-velocities zones with similar widths [13]. Maps obtained for $\sigma$ =[1-10km] are presented in the electronic supplement. Notice that the total weight transfer at this spatial scale ($\sigma$=3km) is merely 0.23% and the maximum weight transfer at each grid cell does not exceed $1.5*10^{-10}$. Due to the omission of the depth component in this 2D illustration, weight transfers on dipping faults are projected on the surface and hence the depletion/enrichment regions close to these faults may be exaggerated, but nevertheless limited at the scale of analysis ($\sigma$). In many places, such as the Brawley and Laguna Salada fault zones in the South-East, the structures highlighted by weight increase coincide with the observed and extruded fault traces. It is possible to infer larger structures such as the San Jacinto Fault and the San Andreas Fault. In other parts of the map, regions of weight accumulation seem to produce patch-like features without a linear structure. This is mostly observed on the San Clemente Fault where condensation highlights small features perpendicular to the fault trace. One explanation could be that this is due to issues of offshore network coverage; however for the Santa Cruz - Santa Catalina Ridge Fault Zone, which is also offshore, we observe a good correspondence. These results, together with the likelihood gains reported in the previous section, suggest that the condensation method can be used to complement fault network reconstruction applications, which rely on high quality location data [2,14].

## 4- Multifractal properties of the original and condensed catalogs

In this section, we investigate the implications of the condensation method for the multifractal properties of the seismicity distribution. Such an investigation is of particular interest because it provides a quantitative description of the spatial patterns at various scales. Since condensation allows us to incorporate the location quality information in the form of a scalar weight, we are interested in whether we find the same values in the appropriate scaling regimes for both original and condensed catalogs. Additionally, we are interested in re-evaluating scaling regimes at small scales that until now have remained concealed due to the effect of location errors [15,16].

### *4.1 The multifractal formalism and classical estimation methods*

Classical (Euclidean) geometry is indispensible for many theoretical and technical applications. However, when used to describe nature, it quickly becomes insufficient. Even simple questions such as "How long is the coast of Britain" [17] become problematic as the answer changes as a function of the observation scale. The formalism of Fractals aims to extract this functional form (i.e recursive scaling regime) and provide insight into the underlying phenomena. The term fractal (monofractal, or homogeneous fractal) implies that the scaling can be quantified by a single exponent, the fractal dimension. Multifractal distributions can be seen as different sets with different scaling properties interwoven altogether. The consequence is that the singularity of the underlying distribution fluctuates from place to place, the complexity of the structure being fully encoded by the multifractal spectrum. A common way to quantify the fractal or multifractal properties of a given set of data points is to calculate its generalized (Renyi) dimensions [18], given as:

$$D_q = \lim_{\varepsilon \to \infty} \frac{\frac{1}{1-q}\log(\sum_i p_i^q)}{\log(\frac{1}{\varepsilon})} \tag{6}$$

where ε is the scale of observation, $p_i(\varepsilon)$ is the fraction of data points (e.g, estimated measure) within box *i* of size ε, *q* is a real-valued moment order and the sum is performed over all boxes covering the data set under investigation. Varying the *q* parameter, $D_q$ characterizes the scaling of the underlying measure within the distribution. Thus, $D_{-\infty}$ and $D_\infty$ respectively correspond to the local scaling of the lowest and highest density areas, i.e. to the weakest and strongest singularities of the distribution. For monofractal sets, $D_q$ is a constant independent of *q*. For multifractal distributions, $D_q$ decreases monotonically with *q*.

The commonly used multifractal analysis methods can be classified into two broad classes, called fixed-size and fixed-mass methods respectively. Fixed-size methods (FSMs) [19,20] estimate $D_q$ via the scaling of the total mass *M* (i.e number of data points) within a constant *r*-sized sphere, as *r* is increases:

$$\log\left\langle M(<r)^{q-1}\right\rangle \approx (q-1)D_q \log(r) \tag{7}$$

Fixed-mass methods (FMMs) estimate $D_q$ via the scaling of the smallest radius *r* to include a fixed mass *m*, as *m* is increases:

$$\begin{aligned}
&\log\left\langle R(<m)^{-(q-1)D_q}\right\rangle \approx -(q-1)\log(m)\\
&\tau(q) = (q-1)D_q\\
&\log\left\langle R(<m)^{-\tau(q)}\right\rangle^{1/\tau(q)} \approx \log(m)^{-1/D_q}
\end{aligned} \tag{8}$$

Several studies report FMMs to be superior to FSMs [21,22]. For a detailed review of both FSMs and FMMs, the reader is referred to [23].

It is important to note that many previous studies have undertaken the task of estimating fractal dimensions for seismic catalogs. However, most of the published results are questionable because I) the used methods are prone to finite size and edge effects that have not be adequately addressed [23]; and II) of lack of benchmarks with synthetic fractals with analytically derivable fractal dimension. For instance, many studies use the correlation integral [24] to estimate the fractal dimension ($D_2$) of hypocenter or epicenter sets [15,25]. The observed scale ranges in such studies are usually quite limited and, due to the inherent finite size and edge effect, it is difficult to quantify the quality of the measure. Phase transitions, observed in the form of change of slopes, are usually identified manually and their attribution to physical dimensions (such as seismogenic layer thickness) or locations errors (horizontal or depth) remains somewhat speculative. Kagan [16] has made notable efforts to facilitate the applicability of the correlation dimension measure by characterizing and correcting for edge effects and location errors. In the same study, Kagan casted doubt on the significance of fractal analysis performed under these conditions and proposed that studying the higher order point configurations might be a better option. Interestingly, Hirabayashi et al [26] reached the same conclusion when they showed that using fixed-mass methods provides more reliable results in multifractal analysis of seismic catalogs. Nevertheless, even with the use of FMMs, dealing with edge-effects remains a problematic task often tackled by introducing scaling limits or data censoring [26,27].

### *4.2 The Barycentric Fixed Mass estimation Method*

To address the problems associated with the commonly used methods, we have previously introduced a new non-parametric method for multifractal analysis [5]. The so-called Barycentric Fixed Mass (BFM) method incorporates two criteria aimed at reducing edge effects, improving precision and decreasing computation time; a) barycentric pivot point selection and b) non-overlapping coverage. As most fixed mass methods, the BFM method has more stable results at small scales, since it avoids sampling empty spaces by extending the measuring scale to the next neighboring point. Figure 7 compares the performance of the BFM method with two commonly used FMM and FSM. The classical methods converge to the true analytical solution only at high $q$ values, however they tend to saturate at the embedding space dimension ($D_E$=2 for this synthetic case) for negative $q$ values. This saturation leads to significant error in the estimation of characteristic dimensions such as $D_0$, $D_1$ and $D_2$ for multifractals with $D_0$ close to $D_E$. This undermines the reliability of the small-scale results obtained with the classical methods and underlines the importance of conducting synthetic benchmark tests.

### *4.3 Multifractal analysis of the Southern Californian Seismicity with robust estimation of the different scaling regimes*

For a given dataset, the $D_q$ and $q$ values are estimated from the moment curves $\log\left\langle R(<m)^{-\tau}\right\rangle^{1/\tau}$ calculated from fixed-mass spheres covering the point distribution. Incrementing the exponent $\tau$ allows sweeping $q$ in the range of interest. Previous studies have found that, for negative $q$ values, the $D_q$ measure becomes unstable due to the inherent undersampling of the emphasized regions [26,28,29]. That is why we focus our attention on dimensions $D_0$ to $D_5$. $D_q$ is estimated from the $\tau(q)$-moment curve's slope, while $q$ is obtained subsequently as $q=1+\tau(q)/D_q$, and we use $\tau(q)$ values of [-3, -2, -1, 0.1, 1, 2, 3, 4, 5, 6]. This allows to roughly cover the interval $q \in [0;5]$. The mass range $m$ is sampled at logarithmically spaced steps rounded to their closest integer value, given by $m_i=m\ 10^{i\alpha}$, with $\alpha$=0.05, where the smallest possible mass value is $m$=2. The curves of averaged radii versus fixed-mass for the different $\tau$ values are given in Figure 8. Since the configuration of covering spheres is stochastic, we can reduce the variance in the curves by repeating the measurement multiple times and averaging the resulting curves (Figure 8). Averaging multiple realizations reduces the variability observed in the large mass ranges.

Both $D_q$ and $q$ are estimated via the local slope for each $\tau$ exponent. For relatively simple multifractals, such as the widely studied growth process of Diffusion Limited Aggregation, the entire $m$ range can be characterized by a single set of slopes (see Figure 1 of [30]). However, in the case of the Southern Californian hypocenter distribution, we observe phase transitions highlighted by changes in these slopes as a function $m$. The visually identifiable breakpoints in Figure 8 mark the transitions from small to medium scales ($R_{S\text{-}M}$) and medium to large scales ($R_{M\text{-}L}$). These are of particular interest as they might provide insight into different characteristic length scales that govern the seismogenic processes. For instance, in their multifractal analysis of fault networks in Saudi Arabia, Ouillon et al. [29] showed that such characteristic length scales might correspond to the rheological stratification of the crust.

The task of identifying the number of observed scaling regimes and their effective ranges can be viewed as an optimization problem. In this setting, we model the curves as a set of discontinuous and piecewise linear functions where each segment is characterized by its slope and intercept. The segments, defined by their breakpoints, are imposed on all the curves simultaneously. The sum of squared errors (SSE) to be minimized is given as:

$$\begin{aligned} SSE &= \sum_{i=1}^{T}\sum_{j=3}^{S}\left(R_{obs}(m) - R_{Mod}(m,i,j)\right)^2 \\ R_{Mod}(m,i,j) &= a_{ij}m + b_{ij} \quad BP_j < m \le BP_{j+1} \end{aligned} \tag{9}$$

where $T$ is the number of curves corresponding to each $\tau$ exponent, $S$ is the number of segments, $R_{Obs}$ and $R_{Mod}$ represent the observed and modeled ordinates of the curves, $a_{ij}$ and $b_{ij}$ are the slope and intercept estimated for segment $i$ of curve $j$ and $BP_j$ is the breakpoint between segments $j$ and $j+1$. Equation (9) implies that the SSE would tend to zero as the number of segments (i.e. the complexity of the model) is increased. Thus, it becomes essential to include a regularizing term that penalizes the goodness of fit for the complexity of the model. In their study, Seidel et al. [31] address a similar problem of investigating the number of linear trends in the global atmospheric temperature record by comparing different models, using the Bayesian Information Criterion (BIC) [32]. In this study, our goal is to represent the continuous multifractal spectra in an interpretable form, rather than identifying the single best model to describe the data. For this purpose, we use an ensemble approach where we consider best fitting models with different number of segments (see Figure 9). Each model is essentially a staircase function defining constant slopes in each segment. By averaging the different staircase functions obtained for each curve, we are able to obtain a continuous set of slopes. These are used in Equation (8) to obtain the continuous multifractal spectra that specifies $D_q$ for any given $q$ and $R$. The proposed approach is much simpler in the sense that it can be applied without the need to account for correlation between data points, which can become problematic for BIC [31], and it can handle the irregular sampling intervals at small $m$ values.

The ensemble is obtained by averaging the slopes of all the best fitting models with 3 to 10 segments. Our choice for the minimum of 3 segments is based on the general shape of the curves and on previous studies reporting the presence of similar numbers of apparent scaling regimes [15,16]. We confirmed that our results are stable with respect to these initial choices by varying the minimum and maximum number of segments.

### *4.4 Multifractal analysis of the Southern Californian Seismicity: condensed versus original catalogs*

The continuous multifractal spectra obtained for the original and condensed Southern Californian seismicity with M≥2 are given in Figure 10. For the original catalog, we observe changes in the scaling regimes occurring at several scales. The first scale is $R_{M\text{-}L}$≈10km ($m$≈400): this scale can be inferred as the effective thickness of the crustal seismogenic width (2R≈10-25km [33]). The decrease of dimensions at this scale can be understood by considering the case of a plate with finite thickness $a$ that is sampled with a uniform point distribution; spheres with $R<a$ will report $D$=3 while, for $R>a$, the spheres will be insensitive to the thickness and thus report $D$=2. We observe a similar decrease of the dimensions $D_{q>2}$ beginning at $R_{M\text{-}L}$≈10km. This scale is also consistent with the depth distribution of the catalog (mean μ= 7.95km, standard deviation σ= 4.44km; 2$R_{M\text{-}L}$≈20km corresponds to the 0.98 percentile). Another transition between small and medium distances is $R_{S\text{-}M}$≈1.5km ($m$≈6): *a priori* it is difficult to conclude if it stems from a genuine physical process or from location uncertainties. However, if the latter is true, we would rather expect generalized dimensions close to 3 at $r<R_{S\text{-}M}$.

For the sake of clarity, we shall investigate the clustering properties in terms of $D_2$ and the multifractality in terms of $\Delta D=D_1-D_5$. At small scales ($r<R_{S\text{-}M}$), the clustering is more pronounced with low $D_2$ = 0.9, while for medium scales ($R_{S\text{-}M}<r<R_{M\text{-}L}$), it is significantly reduced as evidenced from the high values of $D_2$=1.8-1.9. At large scales ($r>R_{M\text{-}L}$) we observe $D_2$=1.2-1.3. The multifractality is strongest at the small scales with $\Delta D$=0.5, decreasing at medium scales to $\Delta D$=0.3 and decreasing even further to $\Delta D$=0.2 at large scales.

Similarly, for the condensed catalog (Figure 10b), we observe two scaling breaks close to the ones reported above. The gradual transition at larges scales becomes more pronounced and remains at $R_{M\text{-}L}$≈10km; however, the small to medium scale transition is shifted from $R_{S\text{-}M}$≈1.5km to $R_{S\text{-}M}$≈2.5km. Furthermore, we observe a significant decrease of clustering within the small scale regime ($r<R_{S\text{-}M}$) as $D_2$ increases to 1.2. A similar increase of the fractal dimensions is also observed for the medium scale range ($R_{S\text{-}M}<r<R_{M\text{-}L}$): $D_2$=1.9-2.0. At large scales ($r>R_{M\text{-}L}$), the $D_2$ values remain similar to the values observed in the original catalog. In terms of multifractality, we observe a significant increase at small scales with $\Delta D$=0.9, while the values at medium and large scales remain similar.

The difference observed in the fractal dimensions of the original and condensed catalogs indicates that the location uncertainty information, which is the basis of condensation, should be an important factor in the spatial clustering analysis of any seismic catalog. As a reminder, the only cause for the different results in the multifractal analyses is the consideration of location error information. Both catalogs have the same event locations and the BFM analyses are conducted in the exact same ways. Previous studies have disregarded location uncertainties in the analysis step only to introduce it to help the interpretation of the scales where the phase transitions occur [15]. It is important to note that, for both the condensed and the original catalog, the small to medium phase transition occurs at distances that are much larger compared to the vertical and horizontal errors of the catalog (see Figure 4). To verify that the observed scaling breaks are not due to location errors, we perturbed the events in the catalog according to their confidence ellipsoids. We repeated the analysis by scaling up the ellipsoids with a factor of 10 and observed that $R_{S\text{-}M}$ increased only by a factor of about 1.5, $R_{M\text{-}L}$ remains unaffected while $D_q$ values on all scales increased. This indicates that the observed $R_{S\text{-}M}$ cannot be due to location errors since one would expect a higher degree of dependency between the two.

### *4.5 Multifractal analysis of a multiscale synthetic dataset*

The results suggest that the Southern Californian seismicity catalog features a distinctive scaling regime at small scales. In order to demonstrate that the methods used in this study are indeed capable of correctly detecting phase transitions and the respective scale at which they occur, we conduct the same analysis on a synthetic dataset. Our goal is to create a dataset that is the result of two different scaling regimes effective on different scales. For this purpose, we generate a spatial density distribution by recursive replication of a 2 by 2 density matrix [2 0; 0 1]. After a number of replications, we modify the density matrix, which is now [2 0; 1 1] (see insets of Figure 11c), and continue the replication process to obtain a multiscale fractal. We then sample the resulting spatial density distribution with weighted points. Figure 11 illustrates the whole process and the obtained multifractal spectra. Since we replicate each matrix 4 times, the phase transition occurs when the linear regime extends to a length of $2^4$=16 units. Thus, the corresponding radius of a covering sphere is $R = 16\sqrt{2}/2 = 11$. The method not only detects the phase transition correctly but is also able to estimate the fractal dimensions for the two scaling regimes accurately.

## 5- Discussion

### *5.1 Consequences for the spatial distribution of earthquake loci*

The results of our analysis reveal the multifractal characteristics of hypocenter distributions, which are evidenced by different scaling regimes holding at different scales. An important question arises regarding the origin of these distinct scales. The largest of these scales (approximately 10km), is common to both the original and condensed catalog and can safely be interpreted as the typical thickness of the seismogenic crust in Southern California. Beyond this scale, seismicity becomes a 2D process, while it is a 3D one at smaller scales. Another phase transition is observed at a smaller scale $R_{S\text{-}M}$≈1.5km for the original catalog. By accounting for the location uncertainties via the condensation method, this transition is offset to $R_{S\text{-}M}$≈2.5km. Below this scale, the effect of the condensation process is also to decrease the strength of the clustering. This can be rationalized by the observation that the compression achieved during condensation (about 7.4% for all events with M>2) is performed by assigning a zero weight to repeating events and thus by removing them (and their associated bias). The change in the location of the scaling break $R_{S\text{-}M}$ may then be also partly a consequence of the change of slope of the moment curves at small and medium scales (see Figure 8). In other words, if we have a bilinear curve and change the slope of only one of the segments the intersection point would shift. A possible physical constraint for $R_{S\text{-}M}$ can be the width of the fault gouge zone. However, Sammis and Biegel (1989) found that the particle size distribution within gouge zones is likely to be a power law with exponent 2.6, suggesting a similar fractal dimension for the set of ruptures bounding the grains and blocks. We observe a substantially smaller value $D_2$=1.2, suggesting that seismicity is indeed much more clustered. Interestingly, such a high degree of clustering seems in agreement with the reports of narrow, quasi-linear seismicity streaks along several faults in California [34–36].

A different explanation for the origin of $R_{S\text{-}M}$≈2.5km can be the earthquake relocation process itself. The relocation is based on a double-difference method using cross correlation of events that are initially clustered according to multiple criteria. In their paper [7], the authors report one of these criteria to be a maximum separation distance of 2.5km. They also report that "If fewer than 150 nearest neighbors existed within 2.5 km, we used Delaunay tessellation to add up to 150 more distant events to each cluster". A conclusive analysis would require repeating the relocation procedure varying this arbitrary distance criterion and repeating the multifractal analysis. The transition scale may also be controlled by the initial clustering criterion such as the correlation coefficient threshold. Although such an analysis is beyond the scope of this study, we also note that double-difference methods have been reported to be highly susceptible to biases resulting from velocity structure errors [37]. These can strongly affect the shape and inner structure of the relocated clusters themselves, hence their associated scaling properties.

### *5.2 Consequences for earthquake triggering models*

The spatial distribution of earthquakes plays an important role in understanding their interactions. Previous studies investigating the importance of small earthquakes in triggering have been mostly limited in reporting only the capacity ($D_0$) [38] or correlation dimensions ($D_2$) [25,39] and using these two dimensions interchangeably (which is valid only under the assumption of monofractality). We argue that the reported values certainly feature strong biases resulting from the applied methods. Moreover, from those values, they were able to draw important conclusions about the triggering properties of events with different magnitudes (and assuming that seismicity can be modeled as an Epidemic Type Aftershock-Sequence (ETAS) process). Within the ETAS formalism [40], the number of aftershocks following a magnitude *M* event is assumed to scale as:

$$n(M) \sim 10^{\alpha M} \tag{10}$$

where $\alpha$ is the productivity parameter. For example, Helmstetter (2003) estimates $\alpha$=0.8. The magnitude-frequency distribution of these aftershocks obeys the Gutenberg-Richter law:

$$P(M) \sim 10^{-bM} \tag{11}$$

In the following, we use $b$=1 as evidenced by global and regional analyses [41–44]. Since each individual aftershock can trigger its own aftershocks, the total number of aftershocks triggered collectively by all magnitude *M* events scale as

$$N(M) = n(M)P(M) \sim 10^{(\alpha-b)M} \tag{12}$$

Equation (12) implies that, if $\alpha>b$, the triggering is dominated by the largest earthquakes, while, if $\alpha<b$, then it is controlled by the smallest ones. The latter case would have serious implications for understanding earthquake interactions, and hence advances in earthquake prediction, since the Gutenber-Richter holds up till very small magnitudes [45], meaning that the majority of the small events are below the detection threshold of current seismic networks [46]. Measuring $\alpha$ from a seismic catalog is problematic not only because it involves subjective definitions of time and space windows for aftershocks, but also because of the inherent incompleteness due to missed events following large main shocks. Here, we argue that fractal analysis can be used to obtain more reliable estimates of $\alpha$. To provide a link between $\alpha$ and the fractal dimension(s) of seismicity, we use the empirical observation that the rupture length *L* of a magnitude *M* event is given by [47]:

$$L(M) \sim 10^{\theta M} \tag{13}$$

where regression analyses show that $\theta$ depends on the faulting style [9]. We also notice that, in case of a fractal distribution, the average number of events within a domain of size *L* scales as:

$$n(L) \sim L^{D_2} \tag{14}$$

Combining Equation (13) and (14) yields the scaling of the average number of aftershocks within a domain with the size of the mainshock:

$$n(M) \sim 10^{\theta M D_2} \tag{15}$$

so that we can identify $\alpha = \theta D_2$. Triggering properties can thus be inferred by comparing $\theta D_2$ and *b*.

Using a correlation integral (i.e. a fixed scale approach), Helmstetter et al. [25] estimated $D_2$=1.5 (also reported by Kagan [16]) and $D_2$=1.74 for two different catalogs in Southern California for $0.1 \leq r \leq 5$km. However, they chose to use $D_2 \approx 2$, estimated for inter-event times larger than 1000 days, in order to remove the distortion of the scaling due to the triggering itself. Citing [47], they chose $\theta$=0.5, and concluded that small earthquakes are as important as big ones for triggering, as $0.5D_2 \approx b$. We notice that, without any constraint on inter-event times, they would have obtained $0.5D_2 < b$, so that small events would be predicted to dominate the triggering. We also notice that Helmstetter [38] uses $\theta$=0.5 while [47] give four possible values of $\theta$ ranging from 0.33 to 1 based on theoretical derivations. Similarly Marsan and Lengliné [39] used 6190 M≥3 earthquakes in Southern California to estimate individually $\alpha$=0.6 and $\theta$=0.43. They reported $D_2$=1.17 and reached the conclusion that small earthquakes have a greater effect on triggering.

We argue that the ETAS formulations can be extended by accounting for the multiscale multifractal characteristics presented in this study and hence provide more rigorous inferences about the earthquake interactions process. As Southern California seismicity is largely dominated by strike slip events, we suggest to use the value $\theta$=0.74. This value has been derived from 43 global strike-slip events with rupture lengths $1.3 < L < 432$km [9]. To illustrate the impact of this scaling parameter, we conduct the following exercise: We calculate the predicted rupture lengths for the 1999 Hector Mine and 2014 M6.0 West Napa earthquakes, based on the rupture length observed during the 1992 M7.3 Landers earthquake. We employ $L(M) = 85km \times 10^{\theta(M-7.3)}$ using the three different $\theta$ values discussed above (see Table 1). We note that small $\theta$ values will exaggerate the triggering effect of small magnitude events since they overestimate observed rupture lengths significantly as the magnitude decreases.

| Observed Rupture Length | | Predicted Rupture Length using 1992 M7.3 Landers (85 km*) | | |
|---|---|---|---|---|
| | | $\theta$=0.43 | $\theta$=0.50 | $\theta$=0.74 |
| 1999 M7.1 Hector Mine | 41 km* | 69.7 km | 67.5 km | 60.4 km |
| 2014 M6.0 West Napa | 12 km** | 23.5 km | 19.0 km | 9.3 km |

* http://www.data.scec.org/significant/chron-index.html
**http://www.eqclearinghouse.org/2014-08-24-south-napa/files/2014/08/EERI-Special-Eq-Report-2014-South-Napa-versionOct19web.pdf

Our results indicate that, for distances $r$<10km, the relocated catalog of Southern California exhibits two different scaling regimes with a transition at $R_{S-M} \approx 2.5$km. If this scaling break is due to the parameter choices of the relocation procedure, we conclude that, using $D_2$=1.9-2.0, we get $0.74D_2 \approx 1.44 > b$, so that large events dominate triggering. On the other hand, if this scale is physical, earthquakes with magnitudes $M$<$M$5.7 (approximate rupture length of 2$R$=5km) induce a triggered seismicity with $D_2$=1.2. The inequality $0.74D_2 < b$ then suggests that at this scale the triggering is controlled by the smallest earthquakes. Yet, for larger magnitudes, the largest events dominate triggering. We anyway underline that such scaling breaks are not compatible with the definition of the ETAS model, which does not feature any transition scale, so that we should be cautious when drawing such conclusions.

The implications of our result, specifically the case of $\alpha>b$, has been previously investigated by Sornette and Helmstetter [48] in terms of the branching ratio:

$$n = \frac{kb}{b-\alpha}\left(\frac{1-10^{-(b-\alpha)(m_{\max}-m_0)}}{1-10^{-b(m_{\max}-m_0)}}\right) \tag{16}$$

where $k$ is a normalization constant of the productivity rate given in Equation 10, $m_{max}$ is the maximum earthquake magnitude (in the range of $M$8-$M$9.5 [49], but see improved methods of determination of $m_{max}$ [50–52]) and $m_0$ is the minimum magnitude of an event that can trigger its own aftershocks. The authors showed that the subcritical, stationary behavior of the earthquake process (i.e that aftershocks sequences die out within a finite time length) requires that $n<1$. Furthermore they demonstrate that the case of $\alpha>b$ in which $m_{max}$ is infinite leads to explosive seismicity dynamics, in the form of stochastic finite-time singularity [53]. Such transient dynamics can actually be observed in various aftershock sequences and also instances of accelerated seismicity. For finite $m_{max}$, such explosive dynamics are transient and taper off before any mathematical divergence, when the largest events of the distribution are sampled.

It is important to note that the conclusions of Helmstetter and Marsan, arguing for the importance of small earthquakes, have spawned case studies aiming to quantify this claim in terms of static stress triggering (e.g [54,55]). Such studies rely on the limited number of available focal mechanisms, various assumptions and large uncertainties to compute Coulomb stresses [56]. Due to these limitations, the results are often inconclusive and difficult to generalize. Here, we showed that a purely statistical and robust approach based on empirical laws can provide rigorous answers to such questions. Although our answers depend on the origin of $R_{S-M}$ , this ambiguity can be resolved by relocating the catalog directly from the waveforms using the fully probabilistic approach NonLinLoc, which gives a more realistic representation of the location PDFs [57]. We shall then be able to conclude on the existence and properties of such a scaling break.

## 6- Conclusion

We have introduced a novel condensation method that improves the spatial information content of seismicity catalogs by accounting for the heterogeneity of the reported location qualities. We obtain significant likelihood gains in synthetic datasets perturbed with realistic location uncertainties, and expect the same to hold for natural ones. Qualitative comparison with mapped fault traces in Southern California indicates that condensation highlights active fault structures. The method also reduces the length of the catalogs significantly and allows the location uncertainty information to be taken into account in spatial analyses. Using this information and the state of the art BFM method, we have performed multifractal analyses on the last 20 years of Southern Californian seismicity. Our analysis reveals a phase transition occurring at $R_{S-M}\approx$2.5km, which is most likely due to the relocation procedure rather than a genuine physical process. We use the correlation dimension $D_2$ obtained in our analysis, together with observations on rupture length scaling with magnitude, to make inferences on earthquake triggering models. Contrary to previous studies, our results suggest that large earthquakes dominate the earthquake triggering process. We thus conclude that the limited capability of detecting small magnitude events cannot be used to argue that earthquakes are unpredictable in general.

We envision that the proposed condensation method will become an essential preprocessing tool in the field of seismicity-based fault network reconstruction, which had significant advances in the recent years [2,14,58,59]. These studies employ clustering methods to infer fault structures illuminated by past seismicity. The vast amount of events contained in the seismicity catalogs hinders the large-scale application of these methods because the computation time increases exponentially with the number of data points. For instance, Ouillon and Sornette [14] used 2747 events of the 1986 Mount Lewis sequence while Wang et al. [2] used 3013 aftershocks of the 1992 Landers event. The condensation method reduces the number of data points significantly and should thus allow for faster computations. With the help of these improvements, we were recently able to perform a clustering analysis on the Southern Californian catalog containing ~500,000 events [60]. Condensation also ingrains the information of location uncertainty into the weight of each event, providing an efficient representation of the relative location quality. Furthermore, the weight of each event can be regarded as a multiplier quantifying repeating occurrences at individual locations, which can be inferred as a representation of repeated slip. By equalizing all the individual condensed weights to 1, we can get better insights into the underlying fault structure. This should be of particular interest because individual faults often have regions with different seismicity rates that hinder a holistic clustering inference.

## Acknowledgments

We would like to acknowledge Shyam Nandan for his valuable suggestions and discussions. The Matlab implementation of the condensation method can be downloaded on the following URL: http://www.mathworks.com/matlabcentral/fileexchange/48702. The Southern Californian seismicity catalog used in this study can be downloaded from the following URL: http://www.data.scec.org/research-tools/alt-2011-dd-hauksson-yang-shearer.html

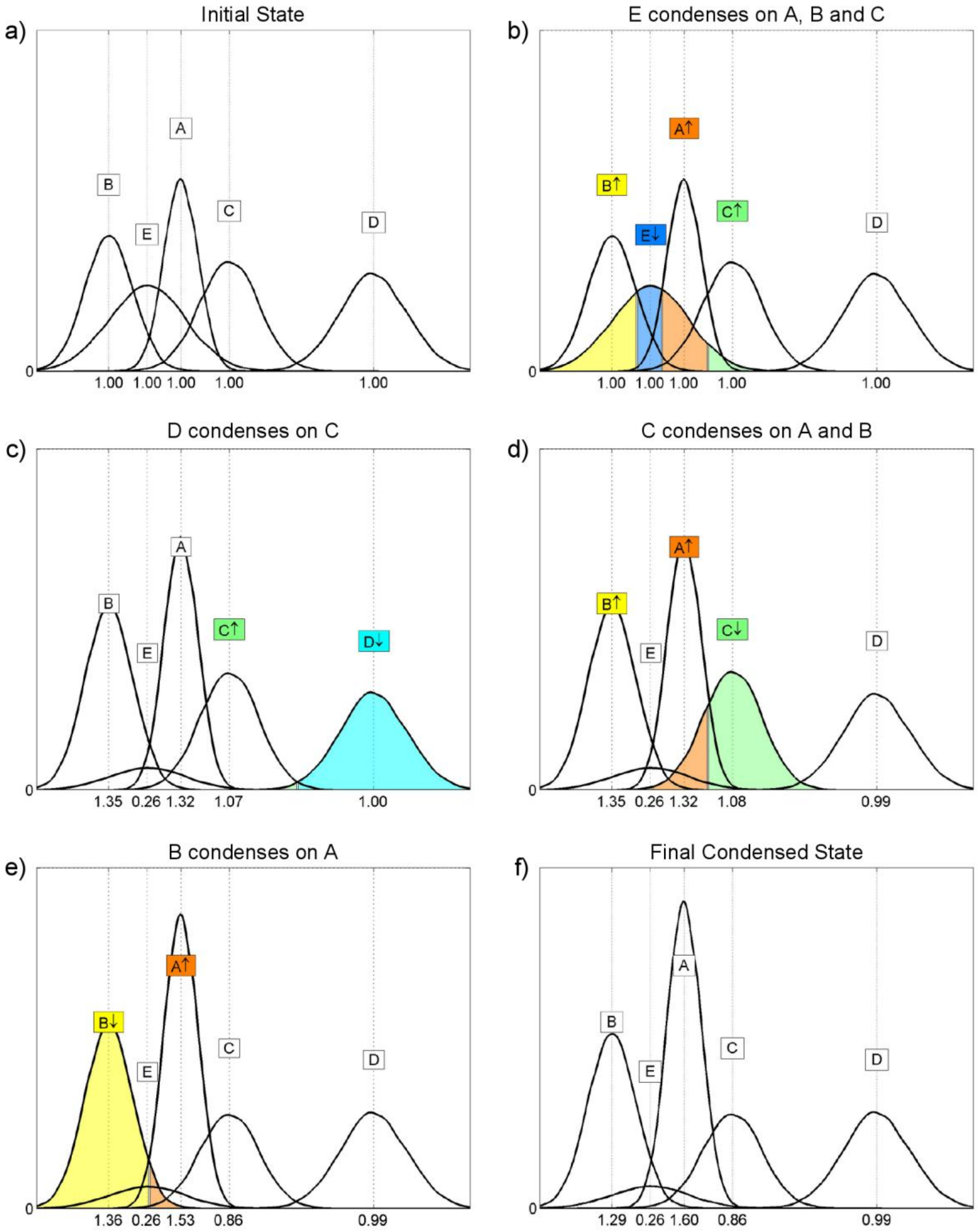

**Figure 1.** Illustration of the condensation procedure for a set of 1D distributions. These are labeled with letters A to E according to their standard deviations [0.5, 1, 1.5, 2, and 2.5]. At each step, source events loosing mass are represented by downward arrows while target events gaining mass are labeled with upward arrows. The portion of the probability weight assigned to each event is depicted with its respective color.

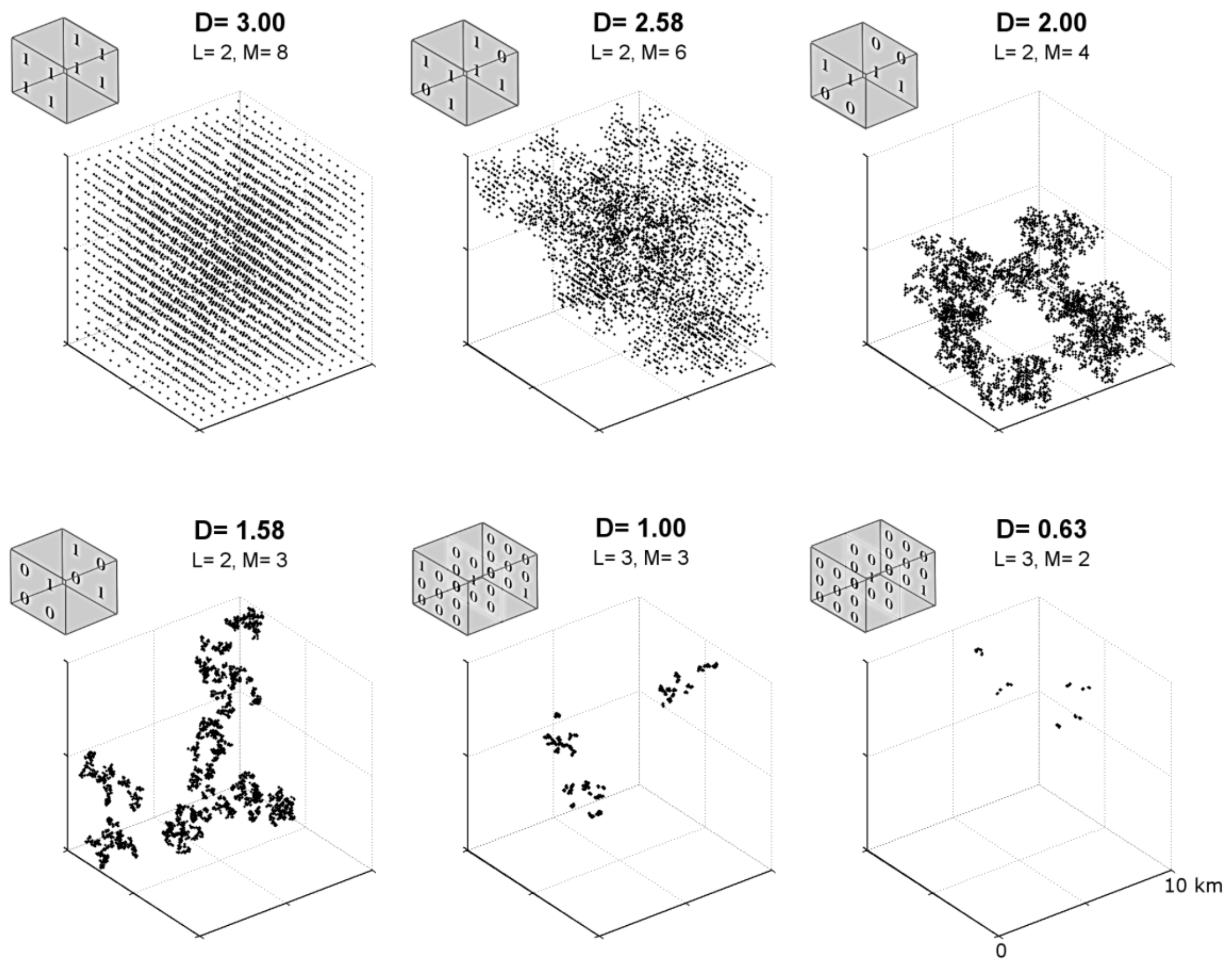

**Figure 2.** Synthetic distributions of 3360 points with different fractal dimension (D). Each distribution is generated by iteratively replicating and permuting the 3D template matrices given in the upper left corners. L and M denote the factors of length reduction and mass increment per iteration, where log(M)/log(L)=D. See Kamer et al. (2013) for details.

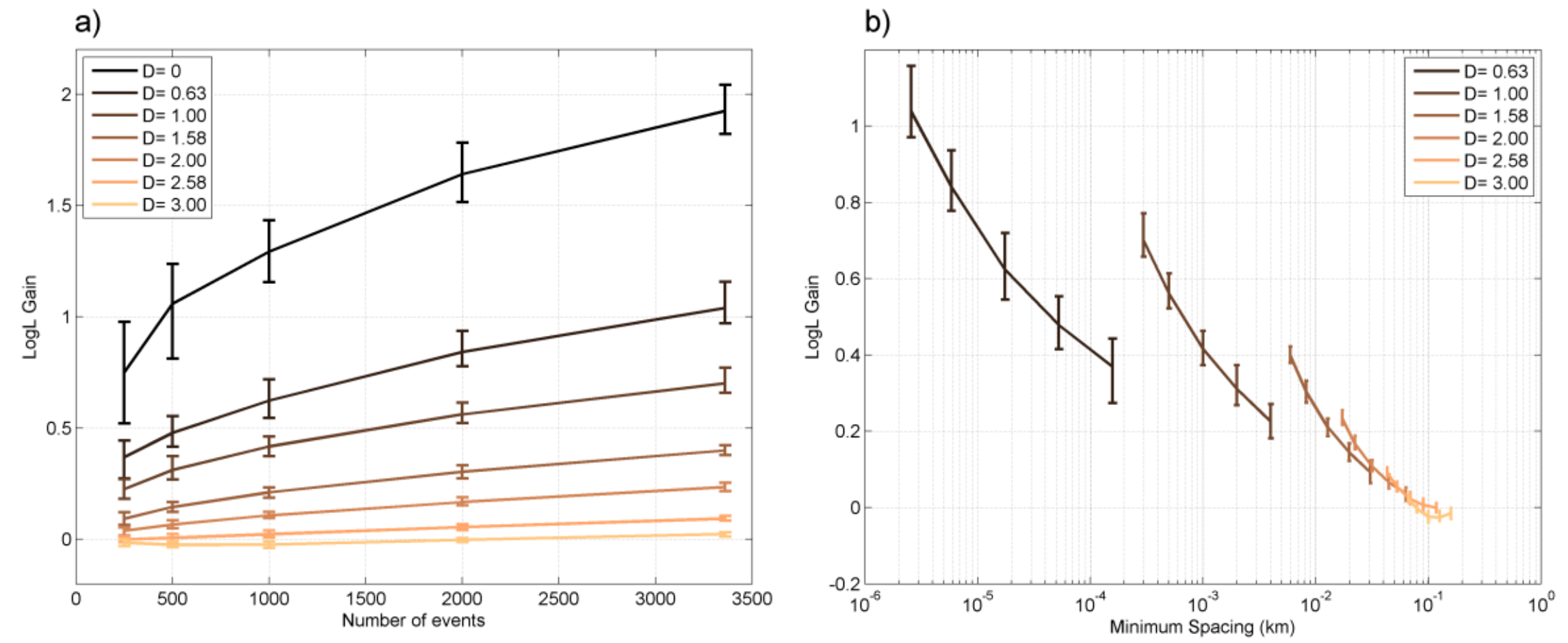

**Figure 3.** a) Log likelihood gain of the condensed catalog with respect to the original catalog. Each curve corresponds to a different distribution with a given fractal dimension D, calculated for an increasing number of events. b) All curves, except D=0, plotted against the minimum spacing calculated from Equation 3.

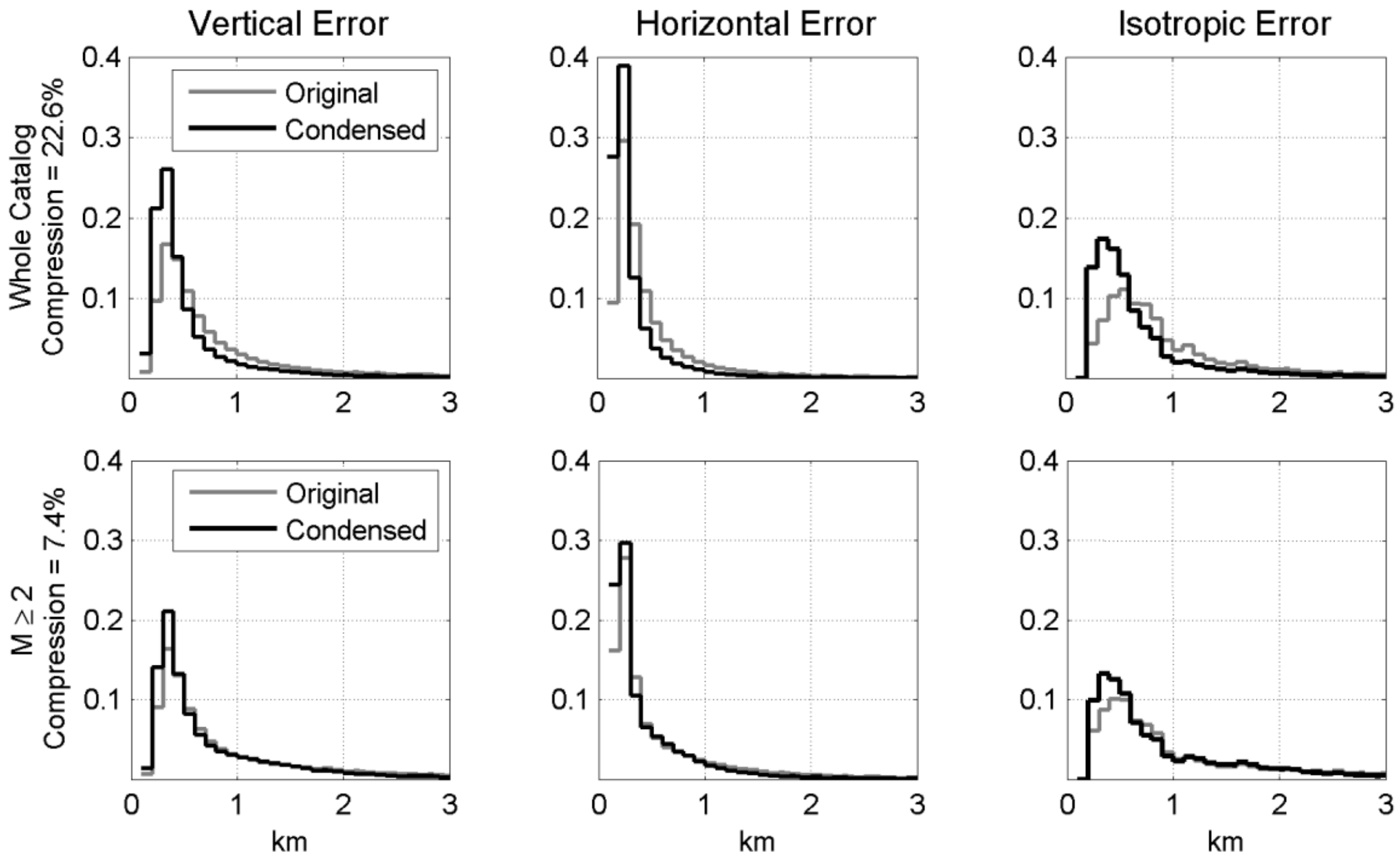

**Figure 4.** Probability density distributions of vertical, horizontal and isotropic errors of the original and condensed catalogs: first row of panels for the whole catalog; second row of panels for events with M≥2

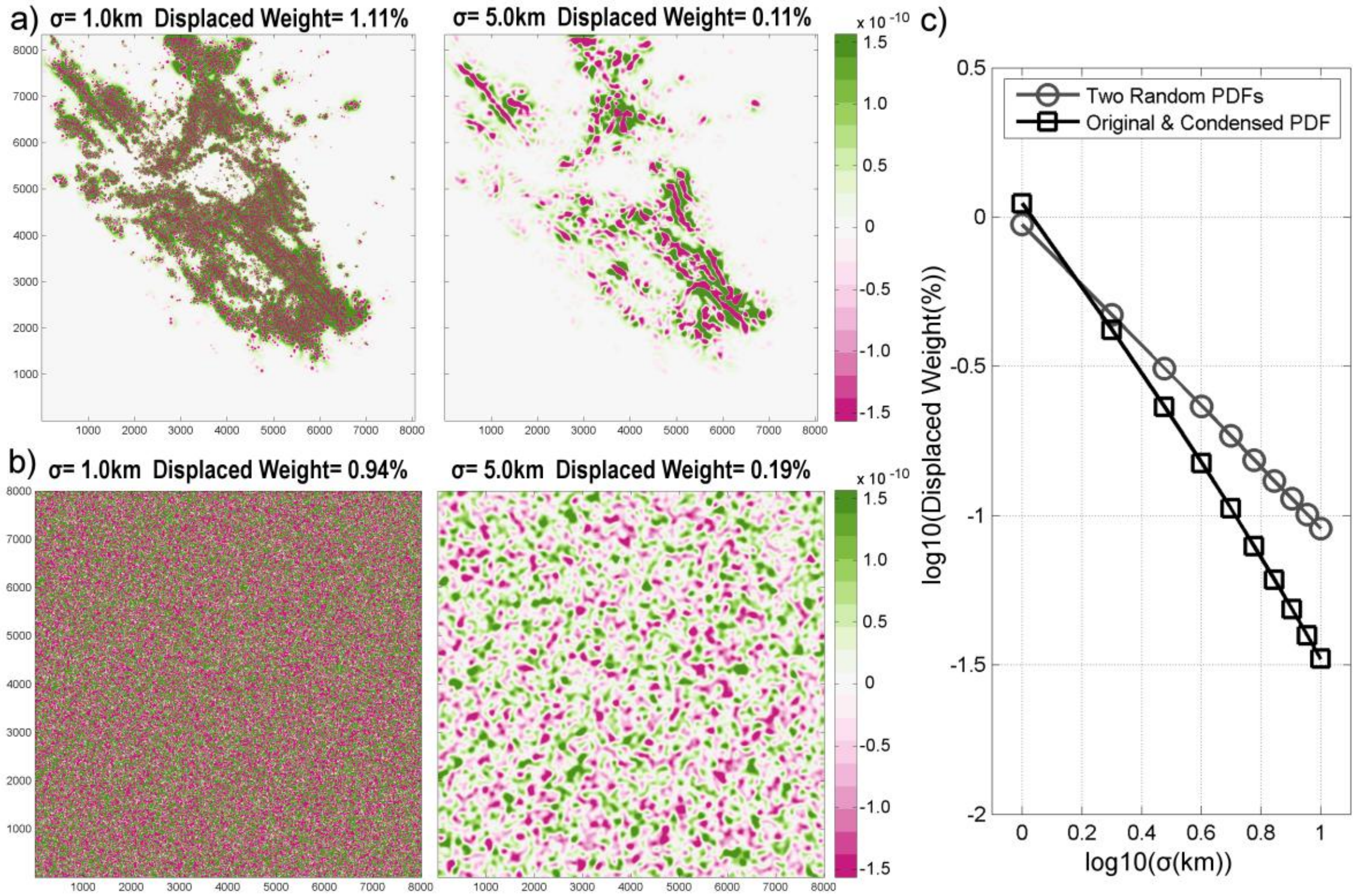

**Figure 5.** Areas of weight enrichment (red) and depletion (green) comparing a) the original and condensed catalogs of Southern California, b) two uniform random spatial PDFs with similar extends for Gaussian filters with bandwidths of σ=1 and σ=5 km. c) Percent of displaced weight as a function of filter bandwidth σ.

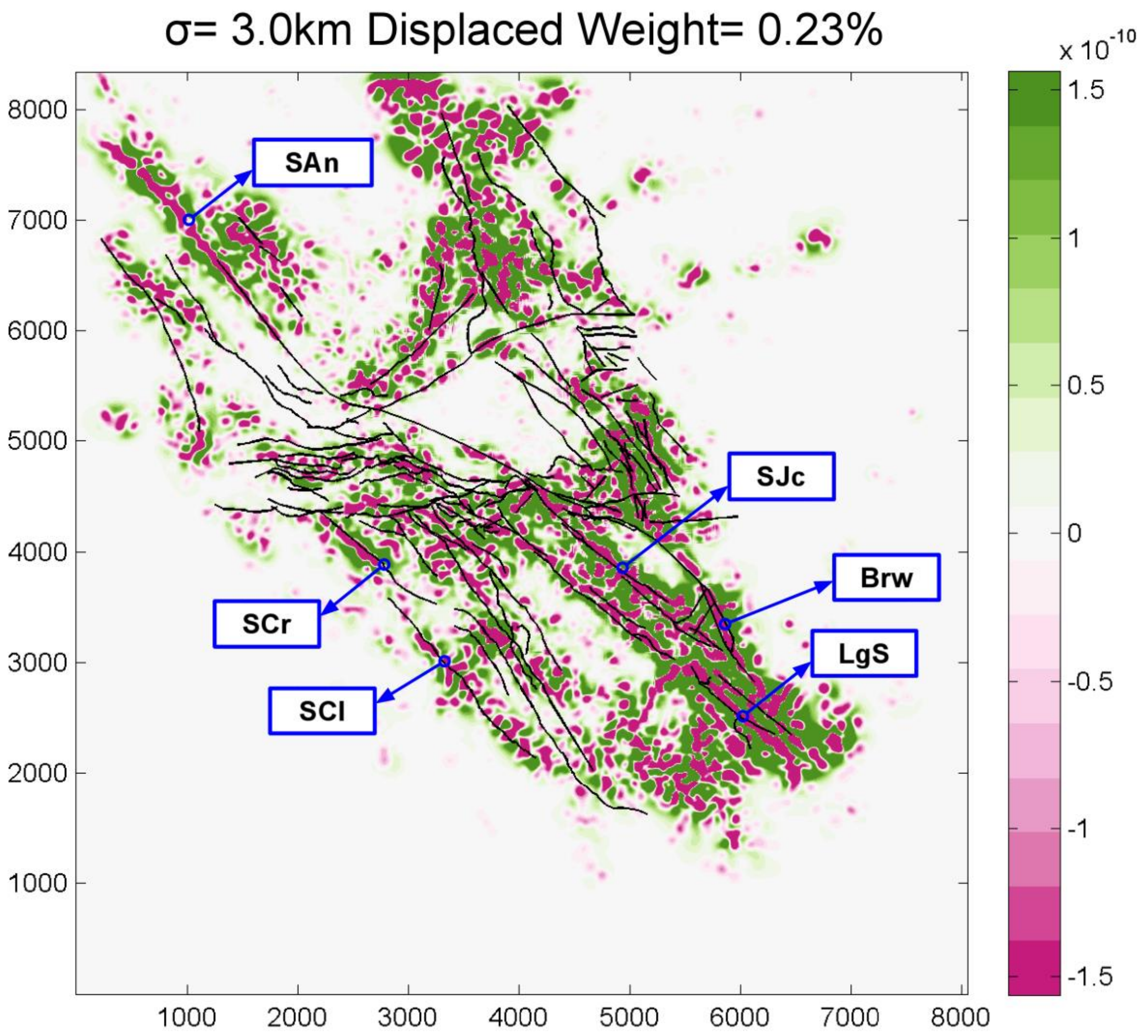

**Figure 6.** Areas of weight enrichment (red) and depletion (green) resulting from condensation at bandwidth of 3km, superimposed with the fault traces obtained from the Community Fault Model. Individual faults are labeled with the following abbreviations: San Andreas (SAn), Santa Cruz (SCr), San Clemente (SCl), San Jacinto (SJc), Brawley (Brw) and Laguna Salada (LgS).

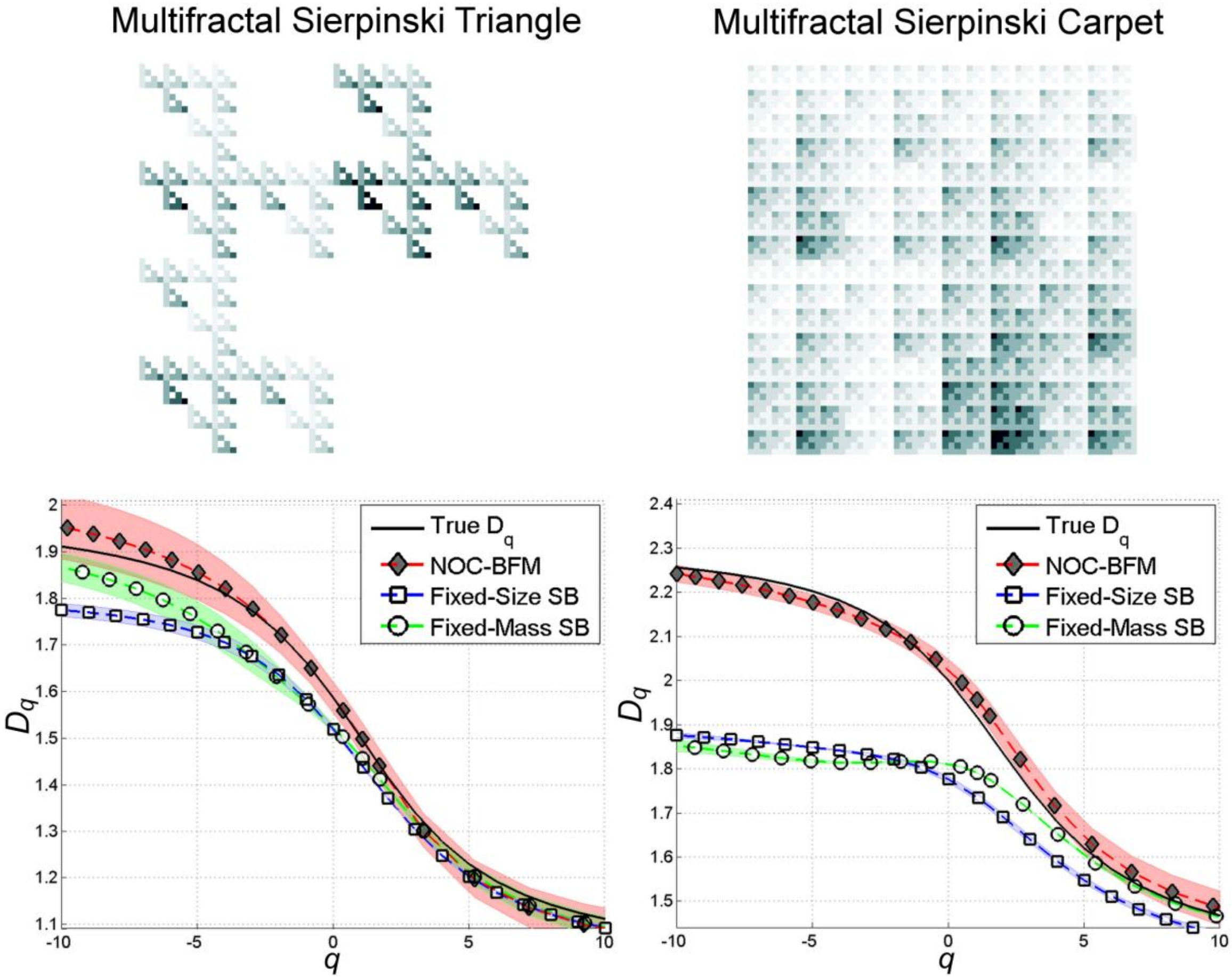

**Figure 7.** $D_q$ vs $q$ curves for two synthetic multifractal point distributions given in the upper right insets: A multifractal Sierpinski triangle (left) and Sierpinski carpet (right). Analytically derived true $D_q$ (black line) is shown together with the BFM method (red), fixed-size (blue) and fixed-mass (green) methods. Reproduced from Figure 6 of *Kamer et al.* [2013]

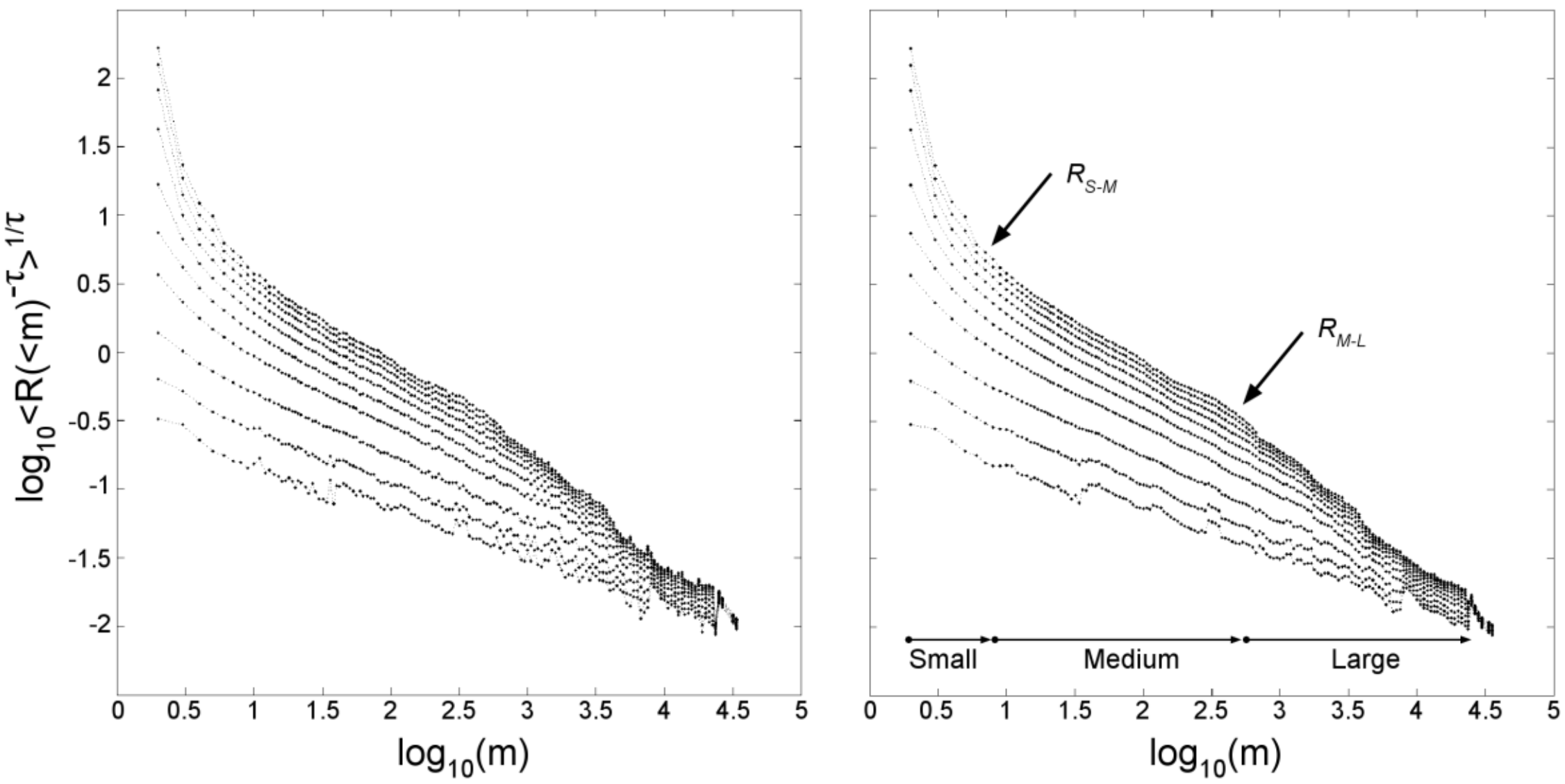

**Figure 8.** Averaged radii versus fixed-mass for increasing $\tau$ obtained from a single measurement (left) and averaged over 100 measurements (right) of the M≥2 Southern Californian seismicity. The two arrows mark the transitions from small to medium scales ($R_{S\text{-}M}$) and medium to large scales ($R_{M\text{-}L}$).

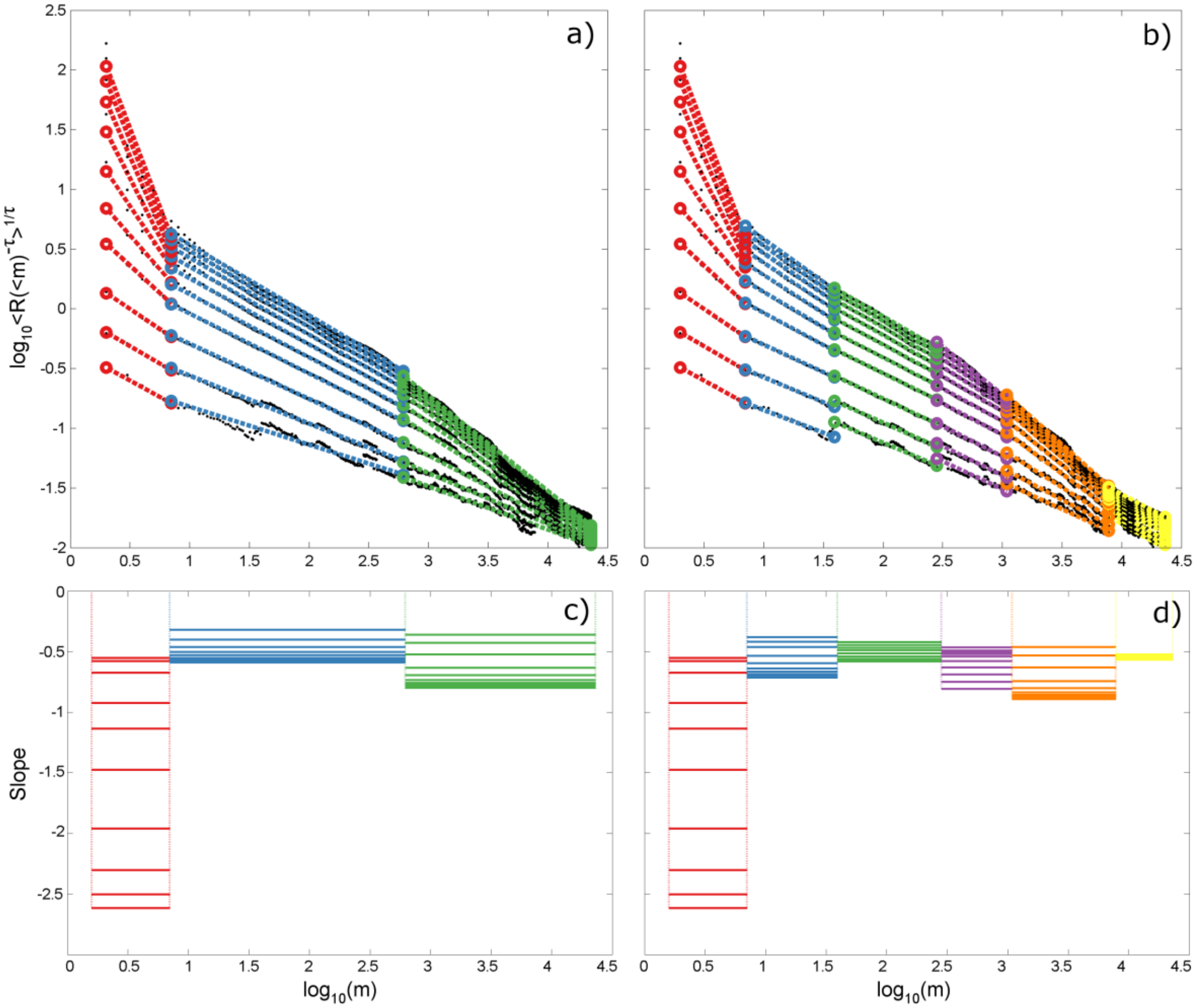

**Figure 9.** Best fitting piecewise linear models with a) 3 and b) 6 segments fitted to the moment curves shown in Figure 8. Different segments are represented by different colors. The staircase slope functions obtained for the best fitting models shown in a,b) are shown in c,d).

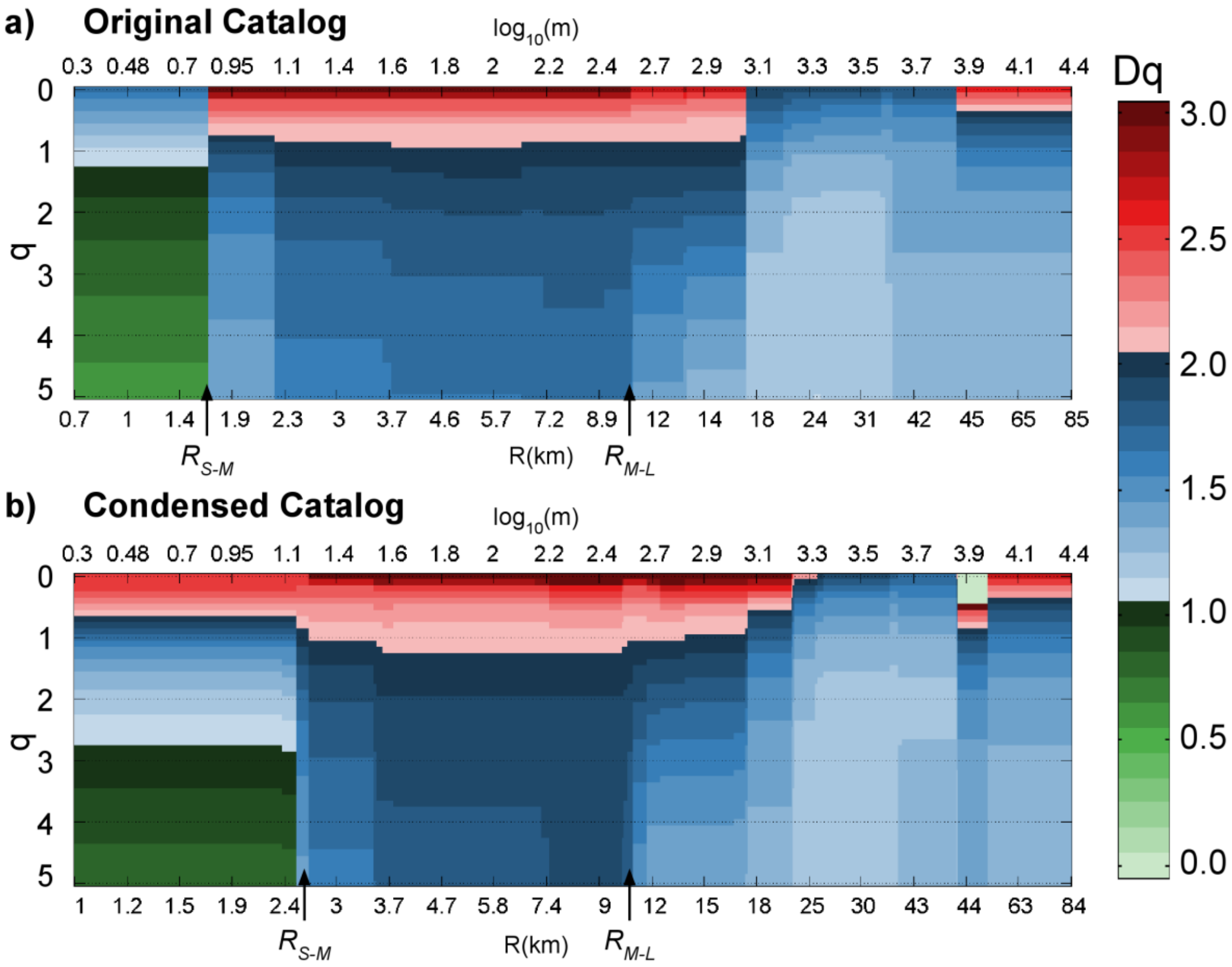

**Figure 10.** Continuous multifractal spectra of the a) original and b) condensed catalogs. The horizontal axis indicates the increasing radius (bottom) and mass (top); the vertical axis represents the $q$ value and individual colors represent $D_q$ within the range [0-3]. The two arrows mark the transitions from small to medium scales ($R_{S\text{-}M}$) and medium to large scales ($R_{M\text{-}L}$).

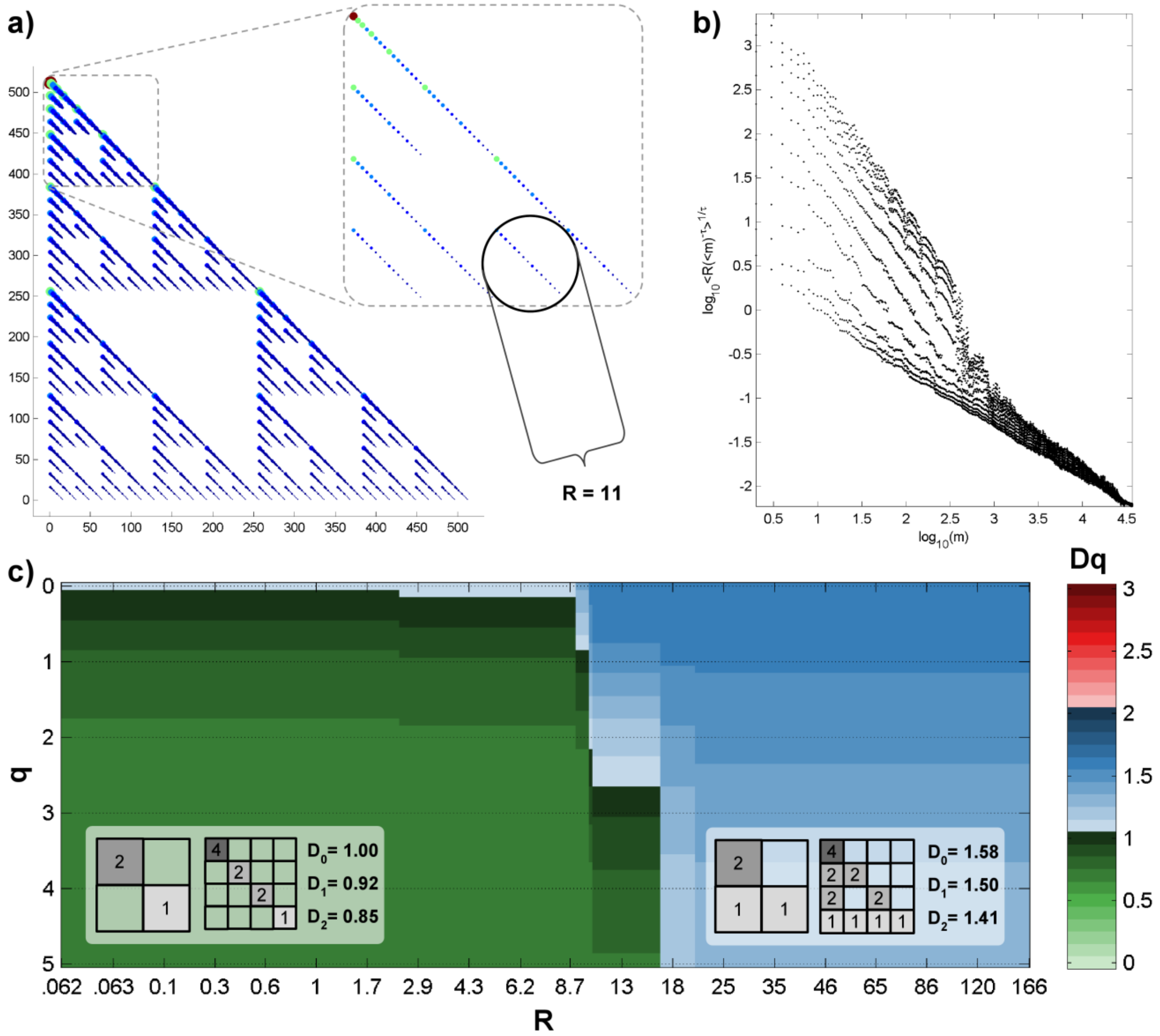

**Figure 11.** a) Spatial distribution of the synthetic distribution, inset shows a close up with the small scale linear features. b) Radii vs mass curves for different increasing τ exponents c) Continuous multifractal spectra of the distribution (similar to Figure 10); insets show the replicating density matrices of each scaling regime together with the analytically calculated $D_0$, $D_1$ and $D_2$ values.